\documentclass[a4paper,12pt]{iopart}

\usepackage{graphicx}
\usepackage{amsfonts,amsthm,bm}
\usepackage{epstopdf}

\newcommand{\CGf}{\ensuremath{\mathcal{C}}}

\newcommand{\RGf}{\ensuremath{\mathcal{R}}}
\newcommand{\ave}[1]{\langle #1 \rangle}

\begin{document}

\title[New transfer-matrix algorithm for exact enumerations]{A new transfer-matrix algorithm for exact enumerations: 
self-avoiding walks on the square lattice}
\author{Iwan Jensen}
\address{\small ARC Centre of Excellence for Mathematics and Statistics of Complex Systems,
Department of Mathematics and Statistics,
The University of Melbourne, VIC 3010, Australia}
\begin{abstract}
We recently published  [J. Phys A: Math. Theor. {\bf 45} 115202 (2012)]
a new and more efficient implementation of a  transfer-matrix
algorithm for exact enumerations of self-avoiding polygons.  Here we extend
this work to the enumeration of self-avoiding walks on the square lattice. 
A detailed comparison with our previous best algorithm shows very significant 
improvement in the running time of the new algorithm. The new algorithm is used
to extend the enumeration of self-avoiding walks to length 79 from the previous record of 71
and for metric properties, such as the average end-to-end distance, from 59 to   71.
\end{abstract}

\section{Introduction}
 
Self-avoiding walks (SAW) on regular lattices is one of the most important 
and classic combinatorial problems in statistical mechanics \cite{Madras93}. 
SAW are often considered in the context of lattice models of polymers 
\cite{Vanderzande98,Janse-van-Renburg00}. 
The fundamental problem is the calculation (up to translation) of the number 
of SAW, $c_n$, with $n$ steps. As most interesting combinatorial problems, 
SAW have exponential growth, $c_n \sim A\mu^n n^{\gamma-1}$, where $\mu$
is the connective constant, $\gamma = 43/32$ is a (known) critical exponent 
\cite{Nienhuis82,Nienhuis84}, and $A$ is a critical amplitude. 
Furthermore the enumeration of SAW have traditionally served as a benchmark for 
both computer performance and algorithm design.
A {\em $n$-step self-avoiding walk} ${\bf \omega}$ on a regular lattice is 
a sequence of {\em distinct} vertices $\omega_0, \omega_1,\ldots , \omega_n$ 
such that each vertex is a nearest neighbour of it predecessor. SAW are
considered distinct up to translations of the starting point $\omega_0$.
We shall use the symbol ${\bf \Omega}_n$ to mean the set of all 
SAW of length $n$. If  $\omega_0$ and   $\omega_n$ are nearest neighbours
we can form self-avoiding polygons (SAP)  by inserting an edge between 
the end-points.

\begin{figure}[ht]
\begin{center}
\includegraphics[scale=0.7]{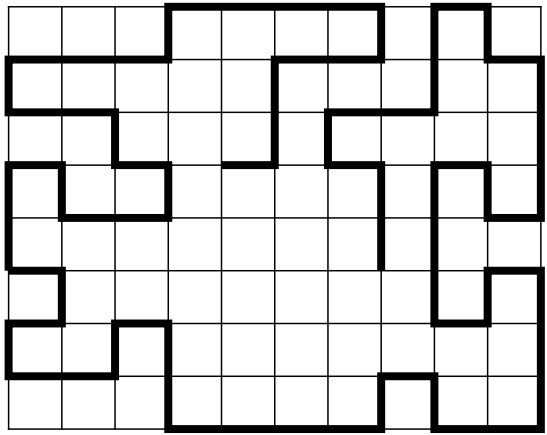}
\end{center}
\caption{\label{fig:sawex}
 An example of a   self-avoiding walk on a $10\times 8$ rectangle.}
\end{figure}

The enumeration of SAW and SAP has a long and glorious history, which for
the square lattice has recently been reviewed in \cite{Guttmann01a}. Suffice to 
say that early calculations were based on various direct counting algorithms
of exponential complexity, with computing time $T(n)$ growing asymptotically 
as $\lambda^n$, where $\lambda = \mu \sim 2.638$, the connective constant 
for SAW. Enting \cite{Enting80} was the first to produce a major breakthrough 
by applying transfer matrix (TM) methods to the enumeration of SAP on finite 
lattices. Results for finite lattices (rectangles in the case of the square lattice) 
are then combined to obtain a series expansion for the infinite lattice.
This so called finite lattice method (FLM) led to a very significant 
reduction in complexity to $3^{n/4}$, so $\lambda = \sqrt[4]{3}=1.316\ldots$. 
More recently we \cite{Jensen99} refined the algorithm using the method of pruning 
and reduced the complexity to $1.2^n$.  Pruning is quite simple in theory and
essentially works by  calculating the minimal order at which a configuration created 
by the transfer matrix  algorithm would contribute to the generations function; 
if this order exceeds a pre-set maximal order we discard (prune) the configuration.
The extension of the FLM to SAW enumeration had to wait until 1993 when
Conway, Enting and Guttmann \cite{Conway93a} implemented an algorithm with 
complexity  $3^{n/4}$.   In 2004 we  \cite{Jensen04}  implemented a
TM algorithm based on the same ideas of pruning used to improve the SAP algorithm.
It appears that this pruning algorithm has a computational complexity of 
$1.334^n$ very close to the CEG algorithm.

All of the above TM algorithms are based on keeping track of the way 
partially constructed SAW are connected to the left of a cut-line bi-secting
the given finite lattice (rectangles in the case of the square lattice). 
Recently Clisby and Jensen \cite{Clisby12} devised
a new and more efficient implementation of the  transfer-matrix
algorithm for SAP.   In that implementation we took a new approach and instead 
kept  track of how a partially constructed SAP must be connected to the right of the
cut-line. In this paper we extend this approach to the enumeration of SAW.
The major gain is that is now quite simple to calculate
the order at which a transfer-matrix configuration will contribute, and this
in turn results in a much faster algorithm.
The draw-back is that some updating rules become much more complicated.

\section{The finite-lattice method and TM algorithms}

All TM algorithms used to enumerate SAW on the square lattice build on the 
pioneering work of Enting \cite{Enting80} who enumerated square lattice 
self-avoiding polygons using the finite lattice method. 
The first terms in the series for the SAW generating 
function can be calculated using transfer matrix techniques to count 
the number of walks in rectangles $W$ unit cells wide and $L$ cells long. 
Due to the symmetry of the square lattice one need only consider rectangles 
with $L \geq W$. Any walk spanning such a rectangle 
has a  length of at least $W+L$ steps. By adding the contributions 
from all rectangles of width $W \leq W_{\rm max}$  (where the choice of 
$W_{\rm max}$ depends on available computational resources) and length 
$W \leq L \leq 2W_{\rm max}-W+1$ (with contributions from 
rectangles with $L>W$ counted twice) the number of walks per vertex of an 
infinite lattice is obtained correctly up to length $N=2W_{\rm max}+1$.

\subsection{Outline of the new TM algorithm}

\begin{figure}[ht]
\begin{center}
\includegraphics[scale=0.9]{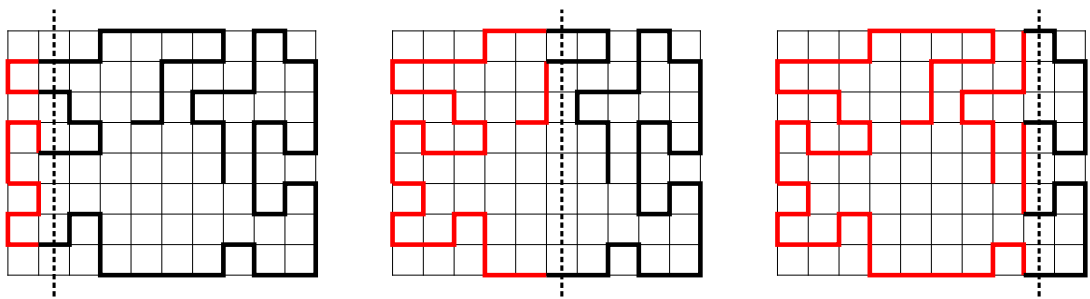}
\end{center}
\caption{\label{fig:sawcut}
 Examples of cut-lines through the SAW of figure~\ref{fig:sawex} such that
 the signature of  the yet to be completed section to the right of the cut-line 
 (black lines) contains, respectively, two, one and no free edges.}
\end{figure}

The basic idea of the new algorithm can best be illustrated by
considering the specific example of a SAW given in figure~\ref{fig:sawex}.
Clearly any SAW is topologically equivalent to a line and therefore
has exactly two end-points. If we cut the SAW by a vertical line as shown 
in figure~\ref{fig:sawcut}  (the dashed lines) we see that the SAW is broken into 
several pieces to the left and right of the cut-line. On {\em either} side of the cut-line 
we have a set of  arcs connecting two edges on the cut-line and at most  
two line pieces connected to the end-points of the SAW.
This means that at any stage a given configuration of occupied edges
along the cut-line can be described in two ways. We can describe how the
edges are connected forming either arcs or line pieces to the left or right of the
cut-line. As we move the cut-line from left to right we can keep track of what happened 
in the past, that is how the pieces are connected to the left, or prescribe what must happen 
in the future, that is how edges are to be connected to the right of the cut-line so as to
form a valid SAW.  The `traditional' TM algorithm keeps track of the past connections.
In the new algorithm we keep track of future connections. There is a natural bijection between 
the configurations which are generated by the two approaches due to the
left-right symmetry of the finite lattice.  The improved performance of
the new approach comes purely from the enhanced efficiency of the
pruning algorithm.

An edge of an arc on the cut-linei is assigned one of two labels depending on whether
it is the lower or upper end of an arc. In addition there are at most two  {\em free} edges 
which are not connected to any occupied edge on the  cut-line. Any configuration along 
the cut-line can thus be represented by a set of edge states $\{\sigma_i\}$, where

\begin{equation}\label{eq:states}
\sigma_i = \left\{ \begin{array}{rl}
0 &\;\;\; \mbox{empty edge}, \\ 
1 &\;\;\; \mbox{lower edge}, \\
2 &\;\;\; \mbox{upper edge}. \\
3 &\;\;\; \mbox{free edge}. \\
\end{array} \right.
\end{equation}
\noindent
If we read from the bottom to the top, the configuration or signature $S$ along the 
cut-lines of the SAW in figure~\ref{fig:sawcut} are, respectively, $S=\{030010230 \}$, 
 $S=\{300000012 \}$, and $S=\{102001002 \}$.
Since crossings are not permitted this encoding uniquely describes 
how the occupied edges  are connected.  As the cut-line is moved a free edge may become
connected  to a new occupied edge thus forming an arc or an existing arc may
form connections to new edges  leading to a `re-configuration' of the arcs.

\begin{figure}[ht]
\begin{center}
\includegraphics[scale=0.7]{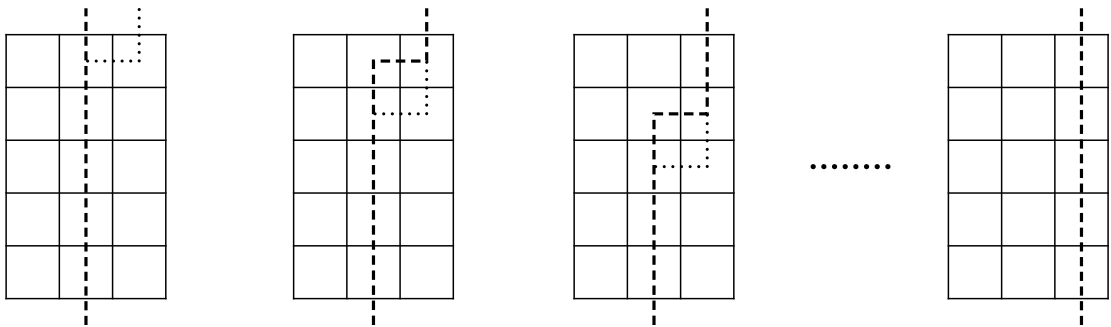}
\end{center}
\caption{\label{fig:transfer}
Snapshots of the cut-line (dashed line) during a transfer matrix
calculation. SAW are enumerated by successive moves of the kink in the cut-line,
as exemplified by the position given by the dotted line, so that vertices
 are added one  at a time.  }
\end{figure}

In applying the transfer matrix technique to the enumeration of SAW
we regard them as sets of edges on the finite lattice with the properties:
\begin{itemize}
\item[(1)] A weight $x$ is associated with each occupied edge.
\item[(2)]  All vertices (except the two end-points) are of degree 0 or 2.
\item[(3)] Apart from isolated sites, the graph has a single connected
component.
\item[(4)] Each graph must span the
rectangle from left to right and from bottom to top.
\end{itemize}
The most efficient implementation of the TM
algorithm generally involves moving the cut-line in such a way as to build up the
lattice vertex by vertex as illustrated in figure~\ref{fig:transfer}.

Constraint (1) is trivial to satisfy. The sum over all contributing
graphs (valid SAW) is calculated as the cut-line is moved through the
lattice. For each configuration of occupied or empty edges along the
intersection we maintain a generating function $G_S$ for partial
walks with signature $S$. In exact enumeration studies $G_S$ is
a truncated polynomial $G_S(x)$, where $x$ is conjugate to the
number of occupied edges. In a TM update each source signature $S$
(before the cut-line is moved) gives rise to new target
signatures $S'$ (after the move of the cut-line) and $k=0, 1$ or 2
new occupied edges are inserted leading to the update
$G_{S'}(x)=G_{S'}(x)+x^kG_S(x)$.

Constraint (2) is easy to satisfy. If both kink edges are empty we can
leave both new edges empty, insert a new arc by occupying 
both of the new edges or we may insert a single new edge. 
If one of the kink edges is occupied then one
or none of the new edges will be occupied. If both of the kink edges are occupied
both of the new edges must be empty. It is easy to see that these rules
leads to graphs satisfying constraint (2). The specific updating rules will
naturally depend on the state of the incoming edges as we shall explain in
some detail below.

In order to satisfy constraint (4) we need to add more information to a
signature. In addition to the usual labeling of the edges intersected by
the cut-line we also have to indicate whether the partially
completed SAW has reached neither, both, the lower, or the upper
borders of the rectangle.  We therefore add two extra `virtual' edge states
$\sigma_b$ and $\sigma_t$ to the signature. Here $\sigma_b$ ($\sigma_t$ ) 
is 0 or 1 if the bottom (top)  of the rectangle has or has not been touched.

We shall now give some details of the updating rules which results in the 
enumeration of all SAW. Constraint (3) will be satisfied by these rules. 
Firstly we look at the construction of the first column of a finite rectangle.
To start a SAW  at a vertex on the left border of the rectangle we insert 
either an arc or a single occupied edge into the totally 
empty configuration. If we insert an arc both occupied edges must be free; 
this is so because there are no other occupied edges along the cut-line 
and  the two edges can't be connected since this would result in a polygon. 
If we insert a single occupied edge it must be free.  Note that it is only during 
the build-up of the first column that this initial insertion into the completely
empty state  occurs and this is the only time that new free edges can be created. 
These rules also ensure that all SAW touch the left-most border of the rectangle. 

The updating rules depends primarily on the states of the two incoming
edges in the kink; secondarily on the state of the edge immediately
below the kink; and finally  edge states  further from the kink
may be affected by the insertion of new occupied edges. The simplest case 
is that in which both incoming edges are occupied.  Recall that the  
encoding in the new algorithm  prescribes how occupied edges are to
be connected.  We can thus join two edges at the kink {\em only} if they belong
to the same arc. Hence the only valid case is the kink-state `12'. Two situations arise 
when arc edges are joined at the kink; {\em either} there are other occupied edges 
along the cut-line and we just proceed with the calculation leaving the outgoing 
edges empty;  {\em or} all other edges are empty and a completed SAW is formed and 
added to the running total for the SAW generating function (provided that the SAW has
touched both the bottom and top border of the rectangle).
All other kink states with two occupied edges (`11', `22', `21', `13', `23', `31', `32', `33') are 
forbidden since they would correspond to connecting occupied edges which 
should not have been connected. Avoiding these situations is part of the
updating rules  for the cases where there is one or two empty edges in the
kink state. Another consequence of this strong restriction on possible
kink states is that we may arrange things such that the vertical kink-edge
is empty unless part of the `12' state. In other words the only possible
kink-states are `00', `10', `20', `30' and `12'.

If the incoming kink state is `10' this edge must be continued
along one of the two outgoing edges. The vertical edge can
only be occupied if the edge immediately below the kink is empty
(otherwise a forbidden kink-state with two occupied edges would be formed).
When the incoming kink state is `20' we continue the walk
along one of the two outgoing edges. The vertical edge can
only be occupied if the edge immediately below the kink is `empty' or `lower'.
In the latter case we check if there are other occupied edges in the
signature; if not we have a completed SAW. When the incoming kink state is `30' we may
continue the walk along one of the two outgoing edges as for the `10' case
or we can terminate the edge at this vertex creating a new end-point for
the walk. In the latter case we again check if there are other occupied edges in the
signature and if not we add the generating function to the running total
as per the `20' case.

\begin{figure}[ht]
\begin{center}
\includegraphics[scale=1.0]{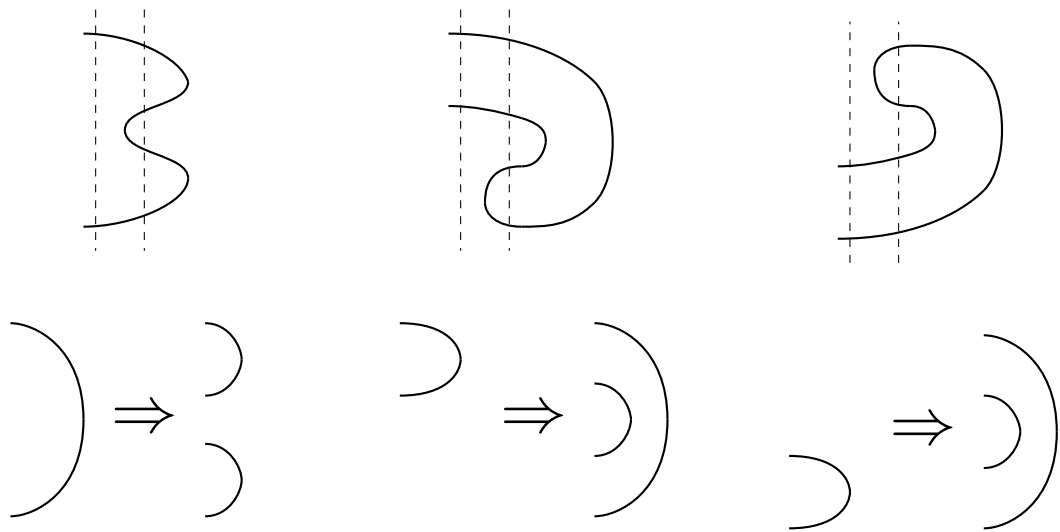}
\caption{\label{fig:basic} 
The possible basic deformations to the topology of a 
configuration with arcs only as the
cut-line is shifted are shown schematically above. The
corresponding basic  updates are shown immediately below.}
\end{center}
\end{figure}

 Finally we turn to the case where the incoming edges are empty. This case
 is by far the most complicated.  Obviously we can leave the outgoing 
 edges empty. If the signature has no free edges the only other possibility
 is to add a new arc on the outgoing edges and connect the new arc to
 an existing arc somewhere else on the cut-line. This update is the
 same as for the SAP algorithm and was described in \cite{Clisby12}.
 For completeness we briefly outline the updating rules for this case.
 The two new occupied edges must connect to existing {\em connected edges} 
 provided these are accessible (more on this later).  In
figure~\ref{fig:basic} we show the two basic situations: The new
occupied edges are either placed inside an existing  arc  
or they are placed outside the arc. In the first
case, shown to the left in figure~\ref{fig:basic}, the upper (lower) end
of the inserted arc must connect with the upper (lower) end of the
existing arc, in terms of the edges involved the states change from
`1002' to `1212' (reading from bottom to top). In the second case, in
the middle of the figure, the upper (lower) end of the inserted arc
must connect with the lower (upper) end of the existing arc, in terms
of the edges involved the states change from `0012' to `1122'. So {\em
both} new occupied edges become `lower' arc-ends while the existing
lower arc-end is changed to an upper arc-end. Shown to the right in
figure~\ref{fig:basic}, there is also a symmetric case where the new
arc is placed above the existing arc leading to the state change
`1200' to `1122'.

\begin{figure}
\begin{center}
\includegraphics[scale=0.7,angle=90]{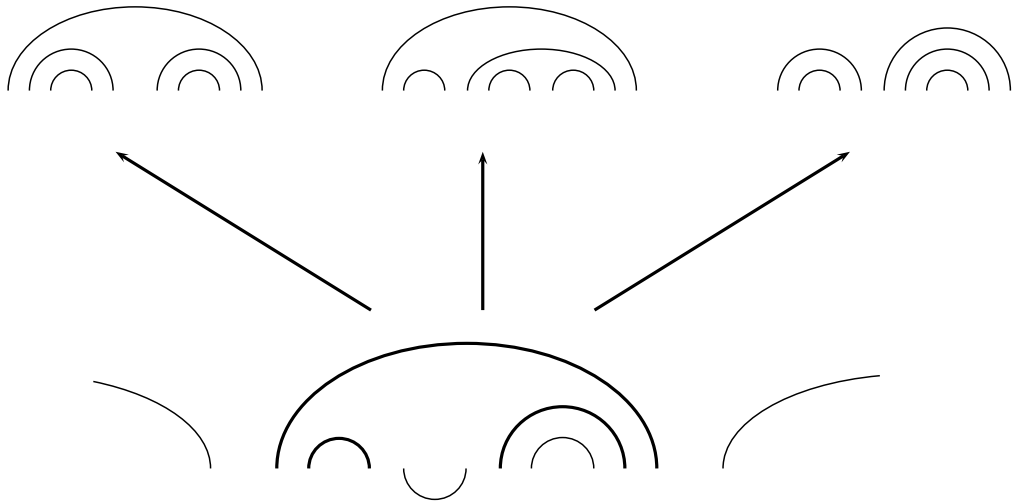}
\caption{\label{fig:insert} 
The possible updates resulting from the insertion of a new partial arc
into an existing arc configuration. At the bottom we indicate by a
lower arc the new partial arc. In the existing arc configuration
(upper arcs) accessible arcs are indicated with heavy lines. The three
possible new arc configurations are shown on top.}
\end{center}
\end{figure}

The newly inserted arc can connect to any existing arc that can be
reached without crossing another arc. The general situation is
illustrated in figure~\ref{fig:insert} where we see that the new arc
can be connected to three existing arcs (indicated by thick lines).
The second arc to the right of the new arc is nested inside an
existing arc and can therefore {\em not} be reached without crossing
the enclosing arc. Likewise any arcs outside the large arc enclosing
the new arc are inaccessible. So in this case the insertion of a single
new arc gives rise to three new arc configurations as illustrated in
the top panels of figure~\ref{fig:insert}. The states of the edges in
the new arc configurations are obtained by applying the appropriate
basic arc insertion from above.

\begin{figure}[ht]
\begin{center}
\includegraphics[scale=0.8]{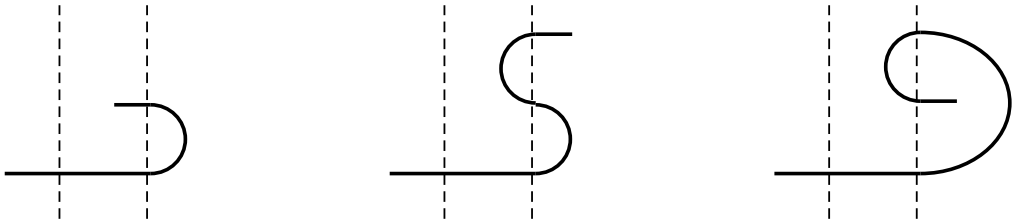}
\caption{\label{fig:basicfree} 
The possible basic deformations to the topology of a 
configuration with a single free edge as the
cut-line is shifted and a single edge or an arc is inserted above the free edge. The
corresponding basic  updates (ignoring empty edges) 
would be `3' becomes `12', `123', and `132', respectively. 
There is a set of symmetric deformations where the insertion happens below the free edge.}
\end{center}
\end{figure}

 Next we turn to the case where there is one or two free edges
 in the source signature.  In figure~\ref{fig:basicfree} we show
 schematically how a configuration involving a single free 
 edge may change as we insert a new occupied edge or a new arc;
 in the latter case the free edge may connect to either the lower or
 upper edge of the new arc.  If there are other  arcs  in the 
 existing configuration these will be unaffected by the insertion. 
 The insertions shown in the figure are possible provided the free
 edge is accessible, i.e., not hidden inside an existing arc. If two
 free edges were present and accessible we could connect the
 newly inserted edge(s) to either of the two free edges and the
 basic update would be the same.
 
 Since we use the form of the TM algorithm where we `grow'
the lattice a single vertex at a time we must split the updating rule  into
 four cases depending on the state of the edge immediately below
 the kink.

 \begin{itemize}
\item[]{\bf Free:} We can insert an arc on the two outgoing edges. The lower
edge of the arc connects to the existing free edge. The local kink
configuration '300' becomes `123'. We can also insert an edge along
one of the outgoing edges. If we occupy the vertical edge this connects
with the free edge below. If no other occupied edges are present a completed
SAW is formed; otherwise we change the 
local configuration `300' to `120'. Finally we may insert a single occupied edge
along the horizontal outgoing edge. This edge must   be connected to
an existing free edge. So we search above (below) the kink for an
accessible free edge and if we find one the state of the outgoing occupied edge 
is `lower' (`upper') while the state of the existing free edge is changed to `upper' (`lower'). 
Note that the insertion of a single occupied edge reduces the number of free
edges by one.
\item[]{\bf Upper/Lower:} We can insert an occupied edge on the horizontal outgoing
edge provided there is an accessible free edge above or below the kink. 
 \item[]{\bf Empty:} In this case we can occupy the horizontal or vertical outgoing edge
with a single edge or occupy both outgoing edges with an arc; in all cases this
only happens if there is an accessible free edge above or below the kink.
In the case where we connect to a free end above the kink the change in the
signature is schematically  that the existing signature $\{000\cdots 3\}$ becomes
 $\{100\cdots 2\}$,  $\{001\cdots 2\}$,  $\{301\cdots 2\}$ and  $\{103\cdots 2\}$, respectively,
 as illustrated in figure~\ref{fig:basicfree}. 
 \end{itemize}

At this stage it seems nothing has been gained. Some updates simplify
while edge- or arc-insertion becomes much more complicated. The   pay-off
comes when we look to pruning. 

\subsection{Pruning}
\label{sec:pruning}

The principle behind pruning is quite simple and briefly works as follows.
Firstly, for each signature we keep track of the current minimum 
number of occupied edges $n_{\rm cur}$ already inserted to the left of the
cut-line in order to build up that particular configuration. Secondly, we 
calculate the minimum number of additional steps $n_{\rm add}$ required to 
produce a valid walk. There are three contributions, namely the number 
of steps required to connect all  occupied edges into a single walk, the number of steps needed (if any) 
to ensure that the walk touches both the lower and upper border, and 
finally the number of steps needed (if any) to extend at least $W$ edges 
in the length-wise direction (remember we only need rectangles
with $L \geq W$). If the sum $n_{\rm cur}+n_{\rm add} > N=2W_{\rm max}+1$ we 
can discard the partial generating function for that configuration,
and of course the configuration itself, because it will not make a 
contribution to the SAW count up to the maximal length we are 
trying to obtain. For instance SAW spanning a rectangle with a width 
close to $W_{\rm max}$ have to be almost staircase like lines, so that
very convoluted  walks aren't possible on these types of rectangles.

In the original approach pruning can be
very complicated. With deeply nested configurations one simply has to
search through all possible ways of connecting existing occupied edges in
order to find the connection pattern, which minimises the number of additional
steps required to form a valid SAW.  
In the new approach almost all of the complications of pruning are
gone. Since connections between edges are already prescribed
there is one and only one way of completing the SAW! The only
complicating factor is that in order to calculate $n_{\rm add} $ we need
to know the nesting level $l$ of each arc. The number of edges
it takes to connect the two edges of an arc at positions $i$ and $j$ is simply
$j-i+2l$. Note that only arcs contribute to the step count required to 
connect the occupied edges in order to form a SAW. Free edges 
only enter into consideration when we have to calculate the number
of steps required to connect the SAW to the lower and upper borders
of the rectangle  and ensure that the SAW extends at least $W$ edges in the length-wise
direction. This pruning procedure can be performed in $O(W)$ operations.

\subsection{Comparative study of the algorithms}

In analysing the complexity of the two algorithms, we note that the
update step when the cut-line is moved may result in $O(1)$ signatures
for the previous algorithm, and $O(W)$ for the new one. However, in the
average case we still expect the new algorithm to create $O(1)$
signatures, as connections with distant occupied edges typically do not
contribute to the generating function (either because the edges are
inaccessible, or the resulting partial SAW configurations have too many
occupied edges and are pruned away).
For pruning, we believe that the complexity of the old algorithm is
exponential in $W$, whereas for the new
algorithm the complexity is $O(W)$.

\begin{figure}[ht]
\begin{center}
\includegraphics[scale=0.5]{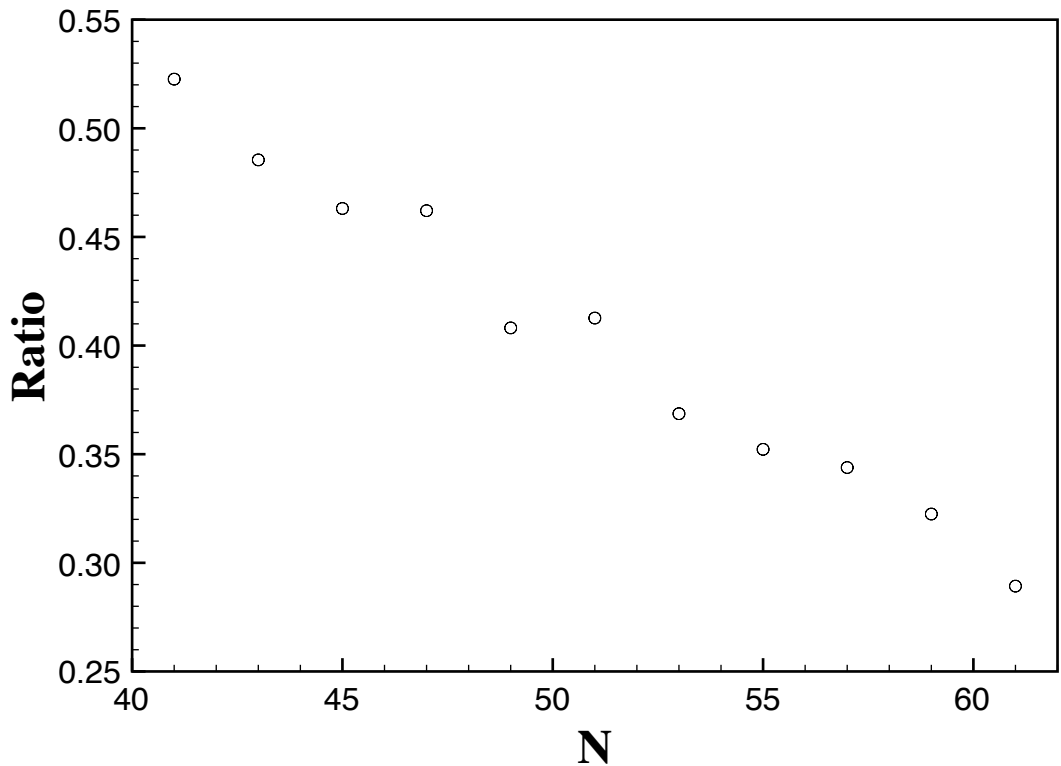}
\caption{\label{fig:time} 
The ratio of CPU time used by the new and old algorithms versus the maximum number
of steps $N$ in the SAW.}
\end{center}
\end{figure}

In table~\ref{tab:algcomp} we compare the resources used by the two
algorithms in a calculation of the number of SAW  of length up to
$N$. From this it is clear that the new approach is much more efficient with
substantial savings in time. The required number of configurations and
terms go down very slightly while the total CPU time decreases by about
70\% for $N=61$.  In figure~\ref{fig:time} we plot the ratio of the time used
by the two algorithms versus $N$ and we notice a clear (reasonably monotonic)
decrease with $N$. The slight scatter in the data can have many causes one
being that the calculations were done on multi-core   processors which can be
sensitive to what else is being computed on the  processor. 

\begin{table}[htdp]
\caption{\label{tab:algcomp} A comparison of the resources required by the
two algorithms in order to calculate the number of SAP up to length $N$.}
\begin{center}
\begin{tabular}{|r|rrr|rrr|}
\hline 
& \multicolumn{3}{c|}{Old Algorithm} & \multicolumn{3}{c|}{New Algorithm} \\
\hline
 $N$ & Configs & Terms & Time & Configs & Terms & Time  \\
\hline
 41 & 258167 & 949978  & 00:04:03 & 237959 & 237959 & 00:02:07 \\
 43 & 458272 & 1656198  & 00:09:10 & 417066 & 1514488 & 00:04:27 \\
 45 & 794580 & 2922585 & 00:20:33 & 728771 & 2669337 & 00:09:31 \\
 47 & 1404805 & 5204306 & 00:45:03 & 1293429 & 4646527 & 00:20:49  \\ 
 49 & 2479142 & 9155623 & 01:33:09 & 2257207 & 8153021 & 00:38:11 \\
 51 & 4303075 & 16201349 & 03:15:17 & 3932460 & 14608709 & 01:20:35 \\
 53 & 7698939 & 28342605 & 06:55:35 & 6945579 & 25443108 & 02:33:12\\ 
 55 & 13457105 & 49905607 & 15:21:27 & 12198903 &44768207 & 05:24:34 \\
 57 & 23486864 & 89241921 &  32:22:55 & 21475037 &78816673 & 11:08:06 \\
 59 & 42037696 & 154839352 & 67:15:57 & 37409087 &138688285 & 21:41:24 \\
 61 & 72844262 & 275510888 & 138:10:11 & 65787437 & 245737647 & 39:58:11 \\
\hline 
\end{tabular}
\end{center}
\end{table}

\section{Extended SAW enumerations}

The transfer-matrix algorithm is eminently suited to parallel
computations and here we used the approach   described in
\cite{Jensen03,Jensen04} to extend the enumeration of SAW from length
71~\cite{Jensen04} to 79. In addition we have extended the series
for some metric properties (such as the end-to-end distance) from
59 to 71 step SAW. 
One of the main ways of achieving a good parallel algorithm using 
data decomposition is to try to find an invariant under the
operation of the updating rules. That is we seek to find some property
of the configurations along the cut-line which
does not alter in a single iteration.
The algorithm for the enumeration of SAW is quite complicated 
since not all possible configurations occur due to pruning,
and the insertion of new occupied edges can change the state of 
an occupied edge far removed. However, there still is an invariant since any edge not
directly involved in the update cannot change from being 
empty to being occupied and vice versa. That is only the kink edges 
can change their occupation status. This invariant
allows us to parallelise the algorithm in such a way
that we can do the calculation completely independently on each
processor with just two redistributions of the 
data set each time an extra column is added to the lattice.  
 
The bulk of the
calculations for this paper were performed on the cluster of the NCI
National Facility at ANU. The NCI peak facility was a Sun Constellation
Cluster with 1492 nodes in Sun X6275 blades, each containing two
quad-core 2.93GHz Intel Nehalem CPUs with most nodes having 3GB of
memory per core (24GB per node). It took a total of about 16500 CPU
hours to enumerate SAW up to length 79. We used up to 400
processors (or more accurately cores) and up to 1TB of memory. Some
details of resource use are given below.

The integer coefficients occurring in the series expansion become very
large and the calculation was therefore performed using modular
arithmetic and the series was calculated modulo various integers $m_i$
and then reconstructed at the end using the Chinese remainder theorem.
In the calculation of $c_n$ we used the moduli $m_0=2^{62}$ and $m_1=2^{62}-1$ 
which allowed us to represent $c_n$ correctly.  The NCI cluster is a
heavily used shared computing facility so our major constraint was CPU
time rather than memory. For this reason we chose to perform the
calculation for all $m_i$ in the same run. Effectively this  increased the
memory requirement by some 50\% but only results in an increase in running time of
some 15\% (compared to a run using a single $m_i$). 

The algorithm for the calculation of metric properties was described
in \cite{Jensen04} and we won't repeat it here. Suffice to say that the 
algorithm requires the multiplication of integers so
to avoid overflow we have to use moduli smaller than $2^{31}$ (signed
integers) and in practise we use the largest prime numbers smaller than $2^{30}$.
In this case we had to use 4 primes to represent the integer coefficients and again we
calculate series  modulo all 4 primes in a single run. The calculation of the
metric properties took a total of about 8000 CPU hours.

Table~\ref{tab:series} lists the new terms obtained
in this work for the number of SAW with perimeter 72--79.  The full series are
available at {\tt www.ms.unimelb.edu.au/\~{}iwan}.

\begin{table}
\caption{\label{tab:series} The number, $c_n$, of embeddings of 
$n$-step self-avoiding walks on the square lattice.  }
\begin{center}
\begin{tabular}{ll} \hline \hline
$n$ & $c_n$ \\ \hline 
72  &  11107224538074654820152678182884 \\
73  &  29442884996760677051402398150644 \\
74   & 78023796077779727644807609460228 \\
75   & 206797849568186990141402577046860 \\
76   & 547952781764285893561169365957068 \\
77   & 1452142167241575828091155500636684 \\
78   & 3847327231644550282490410907667972 \\
79   & 10194710293557466193787900071923676 \\
\hline \hline
\end{tabular}
\end{center}
\end{table}

\subsection{Resource use}

\begin{table}[htdp]
\caption{\label{tab:para} The resources used to calculate the number of SAW on rectangles
of width $W$. Listed from left to right are the number of processors, the total CPU time in hours,
the minimal and maximal number of configurations and series terms retained and finally the
minimal and maximal time (in seconds) used in the redistribution. The minimum and maximum
is taken across all of the processors.}
\begin{center}
\small
\begin{tabular}{rrrrrrrrr} \hline \hline
$W$ & Procs & Time & Min Conf & Max Conf & Min Term & Max Term & $t$-min & $t$-max \\ \hline
20 & 48 & 71 & 13148840 & 13269033 & 105620097 & 110799361 & 810 & 839 \\
21 & 64 & 130 & 17692356 & 17973627 & 123093203 & 128413924 & 1012 & 1047 \\
22 & 96 & 233 & 19262398 & 19818552 & 117757527 & 124452609 & 1212 & 1269 \\
23 & 128 & 366 & 20979889 & 21455635 & 117135449 & 123820823 & 1303 & 1375 \\
24 & 192 & 548 & 18927269 & 19259207 & 101433367 & 105334639 & 1361 & 1427 \\
25 & 256 & 766 & 18383861 & 18745970 & 97671731 & 100083799 & 1457 & 1530 \\
26 & 256 & 1044 & 23823377 & 24298689 & 117585078 & 121505762 & 1692 & 1768 \\
27 & 400 & 1320 & 18909565 & 19281974 & 86963922 & 90679689 & 1697 & 1792 \\
28 & 400 & 1603 & 22401386 & 22874296 & 95884039 & 98305119 & 1779 & 1876 \\
29 & 400 & 1837 & 25481090 & 25840551 & 100310512 & 102930835 & 1814 & 1919 \\
30 & 400 & 2011 & 26643740 & 27211923 & 95761871 & 98905129 & 1843 & 1949\\
31 & 400 & 2134 & 26570470 & 26955370 & 86419643 & 88567259 & 1797 & 2078 \\ 
32 & 400 & 1997 & 23597167 & 23964367 & 65382639 & 67106763 & 1679 & 1800 \\
33 & 400 & 1367 & 17534151 & 18395643 & 46251198 & 51913381 & 1341 & 1426 \\ 
34 & 280 & 685 & 17266488 & 17639105 & 38259003 & 39771128 & 932 & 1047\\ 
35 & 128 & 231 & 13357320 & 23294968 & 24188367 & 45450559 & 480 & 1357\\ \hline \hline
\end{tabular}
\end{center}
\end{table}

In table~\ref{tab:para} we have listed the main resources used by the
parallel algorithm in order to enumerate SAW up to length 79. For
each width $W$ we first list the number of processors used and the total
CPU time in hours required to complete the calculation for a given
width. One of the main issues in parallel computing is that of load
balancing. That is, we wish to ensure to the greatest extent possible
that the workload is shared equally among all the processors. This
aspect is examined via the numbers in columns 4--9. At any given time
during the calculation each processor handles a subset of the total
number of configurations. For each processor we monitor the maximal
number of configurations and terms retained in the generating functions.
Note that the number of terms listed is per modulo $m_i$; so in total three
times this number is actually stored. The load balancing can be roughly
gauged by looking at the largest (Max Conf) and smallest (Min Conf)
maximal number of configurations handled by individual processors during
the execution of the program. In columns 6 and 7 are listed the largest
(Max Term) and smallest (Min Term) maximal number of terms retained in
the generating functions associated with the subset of configurations.
As can be seen the algorithm is very well balanced. Finally in columns 8
and 9 we have listed the minimal and maximal total time (in seconds)
spent by any processor in the redistribution part of the algorithm and
as can be seen this part of the algorithm takes a total of some 15\% of
the CPU time. Note that most of this time is spent preparing for the
redistribution and processing the data after it has been moved. The
actual time spent in the MPI message passing routines is less than 5\%
of total CPU time.

\section{Series analysis and results}

 The number of SAW of length $n$, believed to have the asymptotic 
behaviour
\begin{equation}\label{eq:sawasymp}
c_n  =  A \mu^n n^{\gamma-1}[1+o(1)],
\end{equation}
where $\mu$ is the connective constant and $\gamma$ is a critical
exponent. We shall also study the associated generating function 
\begin{equation}\label{eq:sawgf}
\CGf (x) = \sum_{n=0}^{\infty} c_n x^n = A\Gamma(\gamma) (1-x\mu)^{-\gamma}[1+o(1)],
\end{equation}
so the generating function has a singularity at $x=x_c=1/\mu$.  As for the three 
metric properties we have for the mean-square end-to-end distance of $n$ step SAW 
 that 
\begin{equation}\label{eq:ee}
\ave{R^2_e}_n = \frac{1}{c_n} \sum_{\bm{\Omega}_n} (\omega_0 - \omega_n)^2 =
 C n^{2\nu}[1+o(1)],
\end{equation}
where $\nu$ is a new critical exponent, with associated generating function
\begin{equation}\label{eq:eegf}
\RGf_e (x) = \sum_{n} c_n \ave{R^2_e}_n x^n = 
              AC\Gamma(\gamma+2\nu)(1-x\mu)^{-(\gamma+2\nu)}[1+o(1)].
\end{equation}
The mean-square radius of gyration of $n$ step SAW is defined as
\begin{equation}\label{eq:rg}
\ave{R^2_g}_n = \frac{1}{c_n} \sum_{\bm{\Omega}_n}\left [ \frac{1}{2(n+1)^2} 
\sum_{i,j=0}^n (\omega_i - \omega_j)^2 \right ]=
 D n^{2\nu}[1+o(1)],
\end{equation}
and has an associated generating function
\begin{equation}\label{eq:rggf}
\fl \quad
\RGf_g (x) = \sum_{n} (n+1)^2 c_n \ave{R^2_g}_n x^n \
           = AD \Gamma(\gamma+2\nu+2)(1-x\mu)^{-(\gamma+2\nu+2)}[1+o(1)],
\end{equation}
where the factors under the sum ensure that the coefficients are integer 
valued. Finally, the mean-square distance of a monomer from the end-points of 
$n$ step SAW  is
 \begin{equation}\label{eq:md}
 \fl \quad
\ave{R^2_m}_n = \frac{1}{c_n} \sum_{\bm{\Omega}_n} \left [ \frac{1}{2(n+1)}
\sum_{i=0}^n \left [(\omega_0-\omega_j)^2+(\omega_n-\omega_j)^2 \right ] \right ]
 = E n^{2\nu}[1+o(1)],
\end{equation}
with  generating function
\begin{equation}\label{eq:mdgf}
\fl \quad
\RGf_m (x) = \sum_{n} (n+1)c_n \ave{R^2_m}_n x^n 
           = AE \Gamma(\gamma+2\nu+1)(1-x\mu)^{-(\gamma+2\nu+1)}[1+o(1)].
\end{equation}

The critical exponents are believed to be universal in that they only depend
on the dimension of the underlying lattice. $\mu$ on the other hand is non-universal.
For SAW in two dimensions the critical exponents $\gamma = 43/32$  and $\nu = 3/4$ 
have been predicted exactly, though  non-rigorously, using Coulomb-gas 
arguments \cite{Nienhuis82,Nienhuis84}.

The asymptotic form (\ref{eq:sawasymp}) for $c_n$ only explicitly gives
the leading contribution. In general one would expect corrections to
scaling so 
\begin{equation}
c_n= A\mu^n n^{\gamma-1}\left [1 + \frac{a_1}{n}+\frac{a_2}{n^2}+\ldots
+ \frac{b_0}{n^{\Delta_1}}+\frac{b_1}{n^{\Delta_1+1}}+\frac{b_2}{n^{\Delta_1+2}}+\ldots
\right]
\end{equation}
In addition to ``analytic'' corrections to scaling of the form $a_k/n^k$,
there are ``non-analytic'' corrections to scaling of the form
$b_k/n^{\Delta_1+k}$, where the correction-to-scaling exponent $\Delta_1$ 
isn't an integer. In fact one would expect a whole sequence of
correction-to-scaling exponents $\Delta_1 < \Delta_2 \ldots$, which
are both universal and also independent of the observable. 
Much effort has been devoted to determining the leading non-analytic 
correction-to-scaling exponent $\Delta_1$ for two-dimensional SAW.
In   \cite{Caracciolo05} we studied the amplitudes and the
correction-to-scaling exponents for two-dimensional SAW,
using a combination of series-extrapolation and Monte Carlo methods.
We enumerated all self-avoiding walks up to 59 steps on the square lattice,
and up to 40 steps on the triangular lattice, measuring the
metric properties mentioned above, and then carried out a detailed
and careful analysis of the data in order to accurately estimate the
amplitudes and correction-to-scaling exponent.   This analysis 
unequivocally confirmed that the data is consistent  with the
exact value $\Delta_1 = 3/2$  as obtained from Coulomb-gas arguments
\cite{Nienhuis82,Nienhuis84}. 

Besides the physical singularity  there is another 
singularity at $x=x_-=-x_c$ \cite{Guttmann78,Barber78} which has a
critical exponent consistent with the exact value $1/2$. 
Given the value for the non-analytic correction-to-scaling exponent  
 the asymptotic form used in \cite{Caracciolo05} was
\begin{eqnarray}
c_n & \sim & \mu^n n^{11/32}[a_0 + a_1/n + a_2/n^{3/2}  
             + a_3/n^2 + a_4/n^{5/2} + \cdots] \nonumber \\
   &+ & (-1)^n \mu^n n^{-3/2}[b_0 + b_1/n + b_2/n^2 + b_3/n^3 + \cdots].
\label{eq:cnasymp}
\end{eqnarray}
There will also be sequences of additional terms arising from higher-order
correction-to-scaling exponents.

To test the singularity structure of the generating functions we used
the numerical method of differential approximants \cite{GuttmannDA}. We
will not describe the method here and refer the interested reader to
\cite{GuttmannDA} for details; and Chapter 8 of \cite{PolygonBook} for
an overview of our use of the method.   In
table~\ref{tab:analysis} we list estimates for the critical point
$x_c$ and exponent $\gamma$.  The estimates were obtained by
averaging values obtained from second and third order differential
approximants. For each order $L$ of the inhomogeneous polynomial we
averaged over those approximants to the series which used at least the
first 71 terms of the series. The quoted error for these estimates reflects the spread
(basically one standard deviation) among the approximants. Note that
these error bounds should {\em not} be viewed as a measure of the true
error as they cannot include possible systematic sources of error. Based
on these estimates we conclude that $x_c = 0.3790522766(5)$ and
$\gamma = 1.343745(5)$. The estimate for $x_c$  is consistent
with though several orders of magnitude less accurate than
the estimate $x_c=0.379052277752(3)$ 
obtained from an analysis of the SAP series \cite{Clisby12};
and due to our conservative error-bars  the estimate for $\gamma$
is just consistent with the exact value $43/32=1.34375$.
  
\begin{table}[htdp]
\caption{\label{tab:analysis} Estimates for the critical point
$x_c$ and exponent $\gamma$ obtained from second and third order
differential approximants to the series for square lattice
SAW generating function. $L$ is the order of the inhomogeneous
polynomial.}
\begin{center}
\small
\begin{tabular}{lllll} \hline \hline
 $L$ & \multicolumn{2}{c}{Second order DA} & 
 \multicolumn{2}{c}{Third order DA} \\ \hline 
 & \multicolumn{1}{c}{$x_c$} & \multicolumn{1}{c}{$\gamma$} & 
 \multicolumn{1}{c}{$x_c$} & \multicolumn{1}{c}{$\gamma$} \\ \hline
 0 &  0.37905227478(90)& 1.3437488(14)& 0.37905227604(95)& 1.3437435(27) \\
 2 & 0.37905227492(70)& 1.3437391(34)& 0.37905227663(33)&  1.3437452(14) \\
 4 & 0.37905227679(32)& 1.3437440(12)& 0.37905227658(38)& 1.3437456(16) \\
 6 & 0.379052276419(97)& 1.34374286(30)& 0.37905227672(32)&  1.3437457(12) \\
 8 & 0.37905227617(35)&  1.34374299(63)& 0.37905227670(22)& 1.34374578(84) \\
 10 & 0.37905227630(17)& 1.34374334(46)& 0.37905227670(28)& 1.34374564(88) \\
 \hline \hline
\end{tabular}
\end{center}
\end{table}

\begin{figure}
\begin{center}
\includegraphics[scale=0.85]{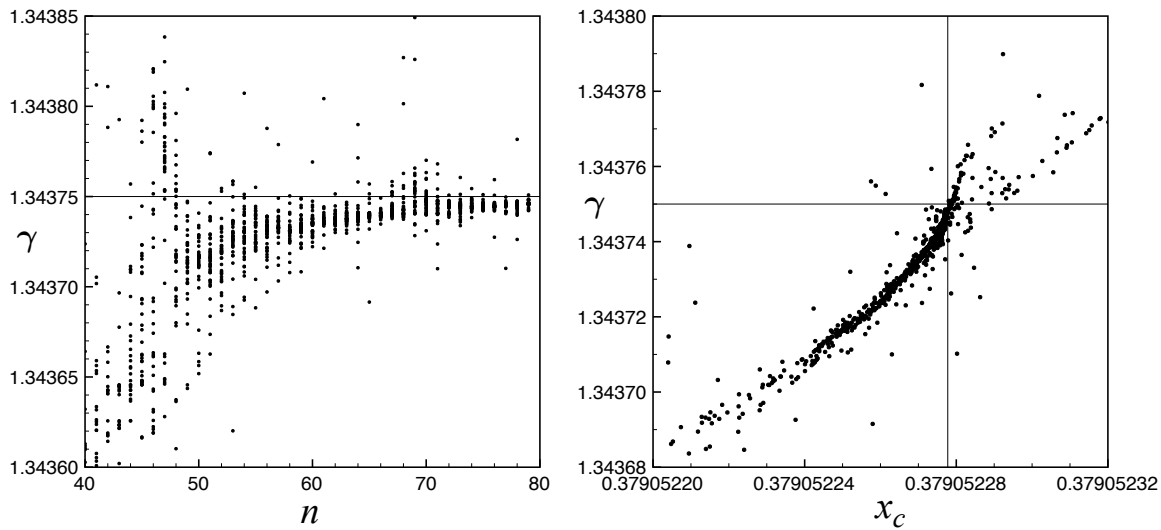}
\end{center}
\caption{\label{fig:crpexp} Estimates of the  critical exponent $\gamma$ versus $n$ (left panel)
 and $\gamma$ versus $x_c$ (right panel) for the
square lattice SAW generating function. The straight lines
correspond to $\gamma=43/32$ and $x_c=0.379052277752$.}
\end{figure}

 To gauge whether or not the estimates truly are as well converged as the results
 in table~\ref{tab:analysis} would suggest we find it useful to plot the actual individual
 estimates against $n$ (where $c_n$ is the last terms used to form a given differential approximant).
 In the left panel of figure~\ref{fig:crpexp} we have plotted the estimates for 
 $\gamma$ as a function of $n$. Each point represents an estimate from
 a third order differential 
 approximant. The approximants appear very well converged and given the very high
 resolution of the abscissa there is no sign of any significant systematic drift. 
 In the right panel we plotted the estimate for $\gamma$
 versus the corresponding
 estimates for $x_c$. If the conjecture for the exact value of $\gamma$ is correct these estimates
 should ideally pass through the point of intersection between the conjectured value and the extremely
 accurate estimate for $x_c$ obtained from the SAP series. All in all these plots clearly are consistent with
 the exact value $\gamma=43/32$.

\subsection{Amplitudes}

In our paper \cite{Caracciolo05} we also obtained accurate amplitude estimates. 
Here we shall therefore only briefly   report on 
the slightly improved estimates for the amplitude based  on directly fitting
our extended series  for $c_n$ to the  asymptotic expansion (\ref{eq:cnasymp}).
For the metric properties we use a similar form except that the
two dominant exponents $11/32$ from the singularity at $x_c$ and 
$-3/2$  from the singularity at $-x_c$ are changed to be 
$59/32$ and $-3/2$ for the end-to-end distance series (\ref{eq:eegf}),
$91/32$ and $1$ for the  mean monomer distance series (\ref{eq:mdgf}), and
$123/32$ and $2$ for the radius of gyration  series (\ref{eq:rggf}).

\begin{figure}
\begin{center}
\includegraphics[scale=0.85]{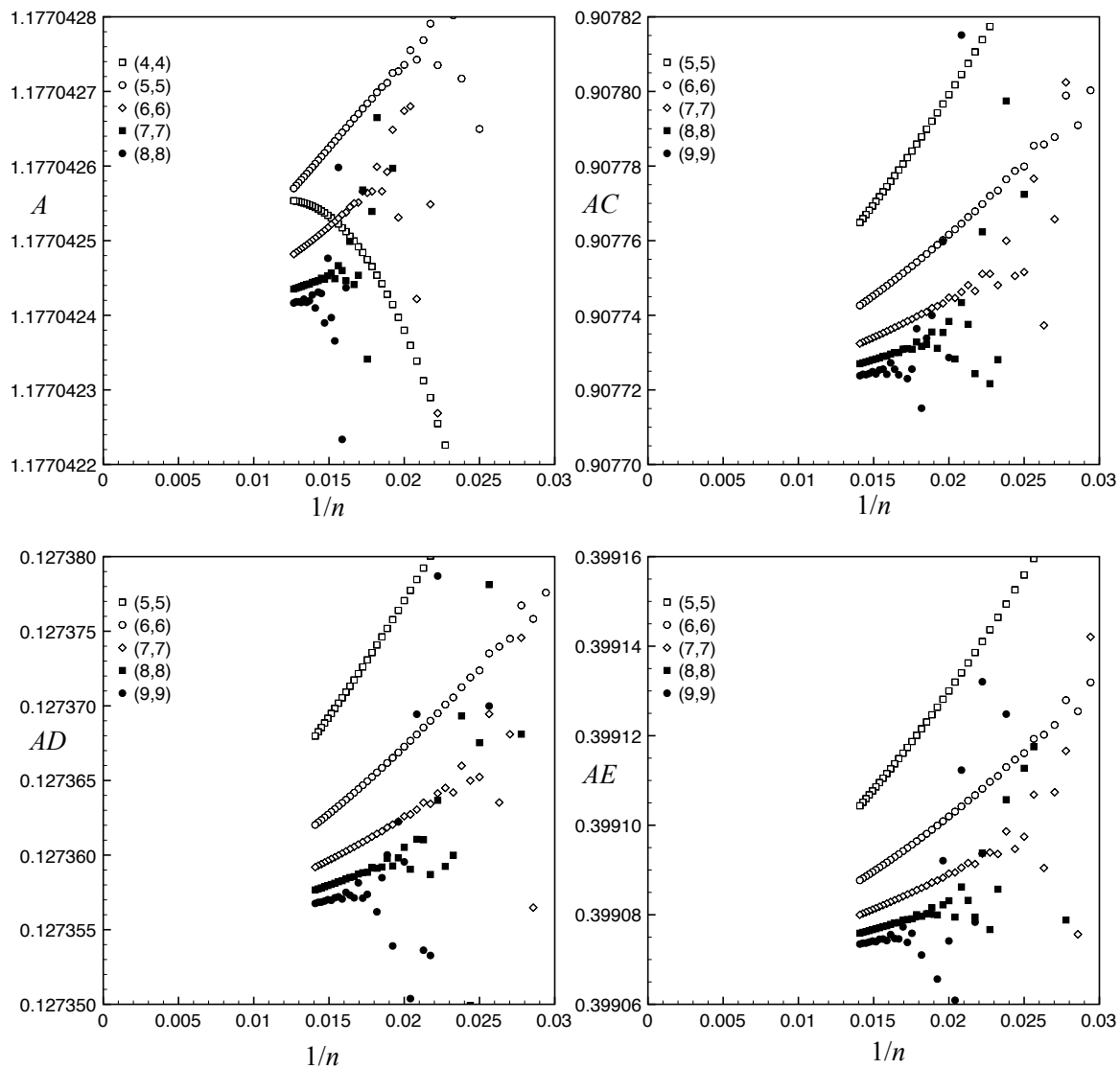}
\caption{ \label{fig:Ampl} 
Estimates for the amplitudes $A$, $AC$, $AD$ and $AE$ versus $1/n$. }
\end{center}
\end{figure}
 
 We find it very useful to plot the amplitude estimates vs. $1/n$ where
$c_n$ is the last coefficient used by the fit. In the top-left panel of 
figure~\ref{fig:Ampl} we plot the estimates for the leading amplitude $A$ from various 
fits. The legend numbers $(k,m)$ indicates the number of terms used
in the fit by each part of the asymptotic expansion (\ref{eq:cnasymp}),
using the exponents given in the  explicit form  (\ref{eq:cnasymp}).
Similar plots for the metric properties are shown in the other panels
obtained from fits using the exponents listed previously. 
From this data we obtain improved estimates for the critical amplitudes for the number of SAW,
 $A=1.17704242(5)$; the end-to-end distance, $C=0.771182(5)$; the radius of gyration, 
 $D=0.1081975(25)$; and the mean monomer distance, $E=0.339043(4)$.

While the amplitudes are non-universal, there are many universal amplitude
ratios. Any ratio of the metric amplitudes, e.g. $D/C$ and $E/C$, is expected 
to be universal \cite{Cardy89}.  
Of particular interest is the linear combination
\cite{Cardy89,Caracciolo90} (which we shall call the CSCPS relation)
\begin{equation} \label{eq:CSCPS}
 F \;\equiv\;
   \left( 2 +  \frac{y_t}{y_h} \right)  \frac{D}{C}
   \,-\, 2 \frac{E}{C} \,+\, \frac12,
\end{equation}
where $y_t = 1/\nu$ and $y_h = 1 + \gamma/(2\nu)$ are the thermal and magnetic 
renormalization-group eigenvalues, respectively, of the $n$-vector model at 
$n=0$. In two dimensions ($y_t = 4/3$ and $y_h = 91/48$, hence 
$2 + y_t/y_h = 246/91$) Cardy and Saleur \cite{Cardy89} (as corrected by 
Caracciolo, Pelissetto and Sokal \cite{Caracciolo90} have predicted, using 
conformal field theory, that $F = 0$. This conclusion has been confirmed by 
previous high-precision Monte Carlo work \cite{Caracciolo90} as well as by series 
extrapolations \cite{Guttmann90,Jensen04}. 
Our new amplitude estimates leads to a high precision confirmation of the
CSCPS relation $F = -0.000006(15)$

\section{Summary}
\label{sec:summary}
 
We have implemented a new algorithm for the enumeration of SAW on the
square lattice; the new method shows considerable promise for future enumeration studies.
The new algorithm was used to extend the series for the number of
SAW on the square lattice from $n=71$ to $n=79$.  Our analysis of the extended
series confirmed that the critical exponents have the exact values $\gamma=43/32$ and $\nu=3/4$.
We obtained improved estimates for the critical amplitudes for the number of SAW,
 $A=1.17704242(5)$; the end-to-end distance, $C=0.771182(5)$; the radius of gyration, 
 $D=0.1081975(25)$; and the mean monomer distance, $E=0.339043(4)$. 
 
 \section*{Acknowledgements}
This work was supported by an award under the Merit Allocation Scheme on the NCI National Facility at the ANU
 and  by  funding under the Australian Research Council's Discovery Projects scheme by the grant
DP120101593.

 \section*{References}

\bibliographystyle{/Users/iwan/Documents/tex/bib/styles/ioptitle_unsrt}
\bibliography{/Users/iwan/Documents/tex/bib/bibfiles/sap,/Users/iwan/Documents/tex/bib/bibfiles/saw,/Users/iwan/Documents/tex/bib/bibfiles/series}

\end{document}